# Recent Progress on Lower Hybrid Current Drive and Implications for ITER

J. Hillairet[1], A. Ekedahl[1,7], M. Goniche[1], Y.S. Bae[2], J. Achard[1], A. Armitano[1], B. Beckett[6], J. Belo[3], G. Berger-By[1], J.M. Bernard[1], E. Corbel[1], L. Delpech[1], J. Decker[1], R. Dumont[1], D. Guilhem[1], G.T. Hoang[1], F. Kazarian[6], H.J. Kim[2], X. Litaudon[1], R. Magne[1], L. Marfisi[1], P. Mollard[1], W. Namkung[4], E. Nilsson[1,7], S. Park[2], Y. Peysson[1], M. Preynas[8], P. K. Sharma[5], M. Prou[1] and the Tore Supra Team[1].

[1] CEA, IRFM, F-13108 St-Paul-Lez-Durance, France.
[2] National Fusion Research Institute, Daejeon, Korea.
[3] Associação Euratom–IST, Universidade Técnica de Lisboa, 1049-001 Lisboa, Portugal.
[4] Dept. of Physics, Pohang Univ. of Science and Technology, Pohang, Korea.
[5] Institute for Plasma Research, Bhat, Gandhinagar, Gujarat, India.
[6] ITER Organization, Saint-Paul-lez-Durance, France.
[7] Department of Applied Physics, Nuclear Engineering, Chalmers University of Technology, SE-41296 Göteborg, Sweden.
[8] Max-Planck-Institut für Plasmaphysik, D-17491 Greifswald, Germany.

*E-mail contact of main author:* julien.hillairet@cea.fr

**Abstract**. The sustainment of steady-state plasmas in tokamaks requires efficient current drive systems. Lower Hybrid Current Drive (LHCD) is currently the most efficient method to generate a continuous additional off-axis toroidal plasma current as well as reduce the poloidal flux consumption during the plasma current ramp-up phase. The operation of the Tore Supra ITER-like LH launcher has demonstrated the capability to couple LH power at ITER-like power densities with very low reflected power during long pulses. In addition, the installation of eight 700kW/CW klystrons at the LH transmitter has allowed increasing the total LH power in long pulse scenarios. However, in order to achieve pure stationary LH sustained plasmas, some R&D are needed to increase the reliability of all the systems and codes, from the RF sources to the plasma scenario prediction. The CEA/IRFM is addressing some of these issues by leading a R&D program towards an ITER LH system and by the validation of an integrated LH modeling suite of codes. In 2011, the RF design of a mode converter has been validated at low power. A 500 kW/5 s RF window is currently under manufacturing and will be tested at high power in 2012 in collaboration with NFRI. All of this work aims to reduce the operational risks associated with the ITER steady-state operations.

## 1. Introduction

The demonstration that a sustainable production of energy is achievable from nuclear fusion in steady-state conditions in a tokamak requires sustaining the plasma current by non-inductive current drive and by the self-generated bootstrap current. The high current drive efficiency of Lower Hybrid waves makes Lower Hybrid Current Drive system a crucial actuator to save inductive flux and to generate additional plasma current in present day tokamak experiments. Several superconducting tokamaks (EAST [1], HT-7 [2], Tore Supra [3], TRIAM-1M [4]) have demonstrated the feasibility of long pulse operation using Lower Hybrid Current Drive (LHCD). Continuous Wave (CW) LHCD capabilities are planned in the KSTAR [5][6] and the SST1 [7] tokamaks. Following the ITER Science and Technology Advisory Committee recommendation, LHCD is considered for a future upgrade of its heating and current drive (H&CD) capability [8].

Simulations of ITER scenarios using LHCD have been updated in references [9] and [10] and show that there is demand for an efficient off-axis current drive. Since this is the main feature of LHCD, LH will reduce the risks associated with the achievement of such steady-state scenarios. Moreover, adding LH power during the current ramp-up phase would efficiently reduce the poloidal flux consumption, resulting in a significant longer flat top for the burn phase [11]. Therefore, preparing the ITER steady state operations not only requires the development of ITER operating scenarios or burning plasma physics, but also to master the LHCD auxiliaries. The achievement of stationary plasmas sustained by LHCD[1] requires addressing a number of technological and physical issues. From the technological point of view, these include the development of high power CW radio-frequency (RF) sources, actively cooled low-loss transmission lines and CW antennas with high coupling efficiency. Real time control systems and reliable predictive coupling and current deposition codes are also required in order to minimize the operational risks associated with the ITER missions. All of these items - from the RF sources to the current deposition scenarios - must have a high reliability and efficiency.

This paper summarizes the recent progress in the LHCD field with respect to the steady-state operations issues listed above and focuses on their potential implications for the ITER project. The paper is organized as follows. The recent achievements in LHCD technologies and operations in the Tore Supra tokamak and LH simulations are reported in section 2. Section 3 emphasizes the potential role of an LHCD system in ITER and reports the current research and development activities conducted by the CEA/IRFM in collaboration with ITER Organization (IO) towards an ITER relevant LHCD system. The LHCD capabilities of the Tore Supra tokamak to address these steady-state issues and the implications to the operational risks reduction in ITER from the know-how and the experiences gained are summarised in the conclusions.

## 2. Recent advances in Lower Hybrid Current Drive

In order to address steady-state operational issues, eighteen 700 kW/CW 3.7 GHz klystrons (Thales Electron Device) manufactured for Tore Supra [12] have been validated on 700 kW/CW matched loads at the in-house test bed facility, before their installation on the LH transmitter [13]. Each klystron produced routinely ~ 620 kW/CW in the Thales Electron Devices test bed, in conditions simulating plasma operation (VSWR =1.4). Eight of them have been installed into the LH plant and operated during the 2010-2011 experimental campaign, generating a coupled power up to 3.8 MW to an actively water-cooled Fully Active Multijunction (FAM) LH launcher [14]. High coupled power, greater than 3 MW for long durations, have been routinely obtained in Tore Supra during the 2011 campaign. A total of 16.65 GJ of accumulated energy has been coupled to the plasma in 2011 by the new klystrons with the FAM launcher.

The second LH antenna of the Tore Supra tokamak is a Passive-Active Mulitjunction (PAM), an ITER relevant concept whose design consists in inserting a passive waveguide (short-circuited waveguide) between two active (directly power fed) waveguides [15]. Such arrangement allows placing water-cooling pipes behind the passive waveguide, thus leading to an efficient exhaust of the thermal fluxes coming from the plasma radiation, RF losses or neutrons damping [16]. Moreover, the passive-active alternation increases the coupling efficiency in conditions with low density in front of the launcher mouth, e.g. at large plasma-launcher gap. Both of these properties solve the general key issues of common LH launchers and are relevant for ITER steady state operations. The PAM launcher has been frequently operated at ITER relevant power density (25 MW/m²) as illustrated in the FIG. 1, in which the power density coupled to the plasma is plotted versus the LH power duration. The capability of the PAM to address long pulse operation issues has been demonstrated in Tore Supra, through the sustainment of a fully non-inductive discharge for more than 50 s, as illustrated in FIG. 2. This launcher is characterized by very low reflected power from the plasma (2% to 6% of the input power) [17], even at a distance from the last closed flux surface (LCFS) of 10 cm in or during fast supersonic molecular beam injections mimicking ELMs in L-mode plasma [18]. A total of 11.4 GJ of accumulated energy has been coupled to the plasma in 2011 with the PAM launcher, which was powered by the old-generation klystrons (400 kW/klystron). Eight 700 kW/CW klystrons have now been installed in 2012 to power the PAM launcher, but have not yet been used on plasma operation.

---

[1] We define here a stationary-state of the plasma as a zero loop voltage discharges over duration much longer than the resistive time scale and the total thermalisation time scale of the tokamak components.

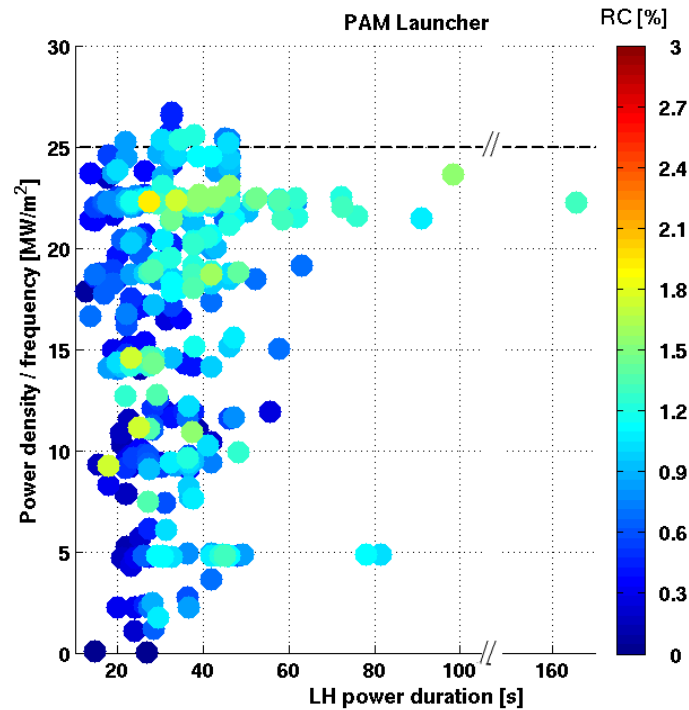

*FIG. 1. Lower Hybrid power density injected into the plasma versus the power duration during the 2011 Tore Supra campaign for the ITER-like PAM launcher. 494 points are plotted, corresponding to the set of the Tore Supra pulses in 2011 having a net power larger than 50 kW. The colour scale illustrates the reflected power (in percentage of the input power) at the rear of the launcher. The dotted line is the power density design value of 25MW/m², which is also the ITER design objective (at 3.7GHz).*

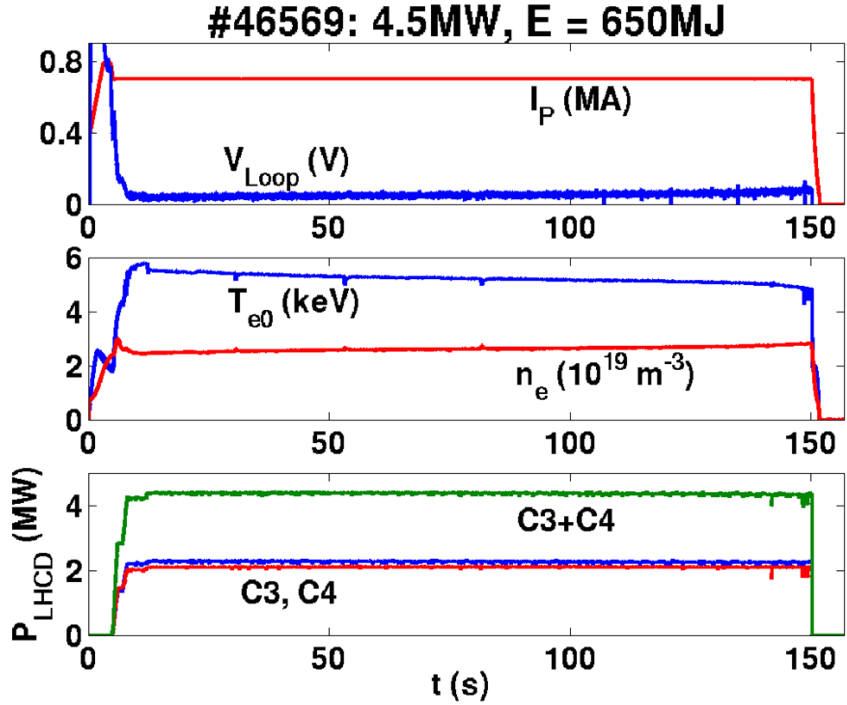

*FIG. 2. Tore Supra pulse #46569 with both the FAM launcher (C3) and the LH ITER-relevant PAM launcher (C4). From Ref.[14].*

Alongside experiments, a general framework is used to interpret Tore Supra LH experiments. This framework includes the LH wave coupling modelling with realistic antennas geometries either in linear regime with the ALOHA code [19] or with the PICCOLO code in non-linear regime [20]. The current deposition properties are calculated by the ray tracing calculations from C3PO associated to the Fokker-Planck solver LUKE [21]. The ray tracing code C3PO takes into account the effect of density fluctuations, and current drive calculations is applicable when power absorption lies far from the fluctuating region. The framework can integrate equilibrium and kinetic profiles from METIS [22] or CRONOS simulations [23], as well as realistic antenna spectra calculated by ALOHA. A fast electron Bremsstrahlung synthetic diagnostic R5X2 reproduces efficiently the HXR signal in Tore Supra steady state discharges at low density, both absolute count rate, HXR profile and photon temperature [24][25][26]. All these tools, which have been validated on low-density plasma ($n_e < 2.10^{19} m^{-3}$), are mature enough to be used with confidence for the design of next-generation LH system.

Alongside the modelling of the LH power deposition in the plasma centre, the unwanted effect of fast electron generation in front of the LH launchers, known to cause hot spots on the lateral protections, has been modelled. From the electric field at the mouth of the launcher, calculated by LH coupling codes, the electron power flow has been evaluated from test electron modelling [27][28] and the associated heat flux on walls has been compared to Infra-Red measurements[29].

### 3. LHCD to sustain steady-state operations in ITER

In the frame of an international collaborative task, coordinated by CEA/IRFM under the EFDA organization [8], the design of the 5 GHz 20 MW/CW LH system for ITER has been updated from the initial 2001 Detailed Description Document in regards of the new results obtained by the LH community. This task, which also involved the training of future RF and mechanical engineers, has now led to the initiation of R&D projects of critical components of a 5 GHz LH system.

The LH system proposed for ITER is made of forty eight identical modules, each one independently fed by one 5 GHz/500 kW/CW klystron (twelve in the toroidal direction and four in the poloidal direction) [31]. The RF power is carried through a transmission line from each klystron up to a double 500 kW RF window located outside the plug frame and connected to a 3 dB splitter which feeds two TE10-TE30 mode converters. After the mode converter, the power is divided into three vertical rows, corresponding to the inputs of a four-active waveguides Passive-Active Multijunction launcher. Thus, a launcher module consists of four active waveguides in the toroidal direction and six (2 × 3) lines of waveguides in the poloidal direction. The whole antenna contains 1152 active waveguides whose dimensions are 9.25×58 mm.

In 2011, a 5 GHz $TE_{10}$-$TE_{30}$ mode converter has been successfully tested at low power at CEA/IRFM and the measurements are in very good agreement with the computations made by the RF modelling [32]. FIG. 3 illustrates the transmission coefficients from the input of the mode converter to its three outputs (S21, S31, S41) and shows a balanced power division (-4.79 dB at 5 GHz, i.e. one-third of the input power). The mode conversion efficiency has been measured by direct electric-probing and found to be more than 99% and with low reflection (-40 dB).

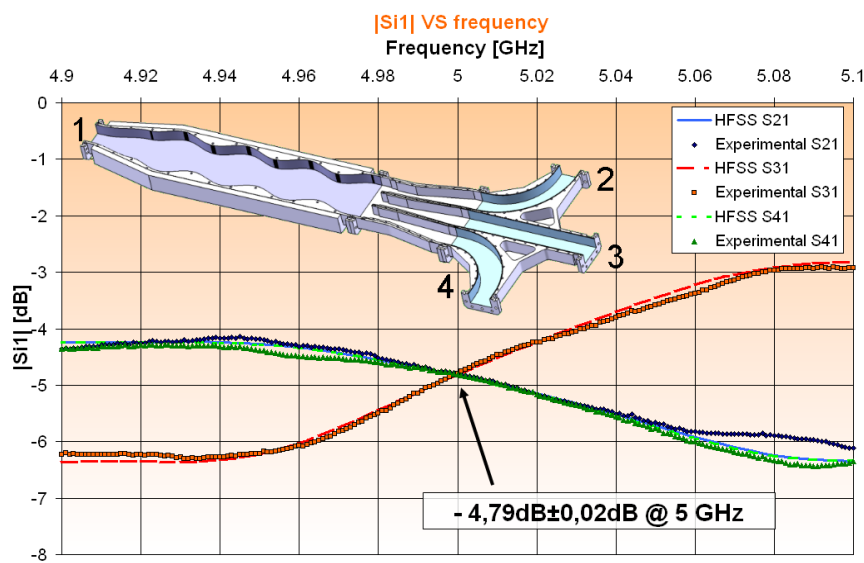

*FIG. 3. Power transmission of the 5 GHz mode converter and power splitter mock-up[32]. The symbols correspond to the three measured transmission coefficients of the mode converter. The lines correspond to the here transmission coefficients calculated by the RF software HFSS.*

In parallel, the space allocation of the LH transmission lines, RF plant and high voltage supply has been performed in association with IO, as illustrated in FIG. 4. Concerning the launcher design, further RF optimizations have been conducted in order to decrease the maximum electric field inside the 5 GHz multijunction and increase the safety margins against possible arc events.

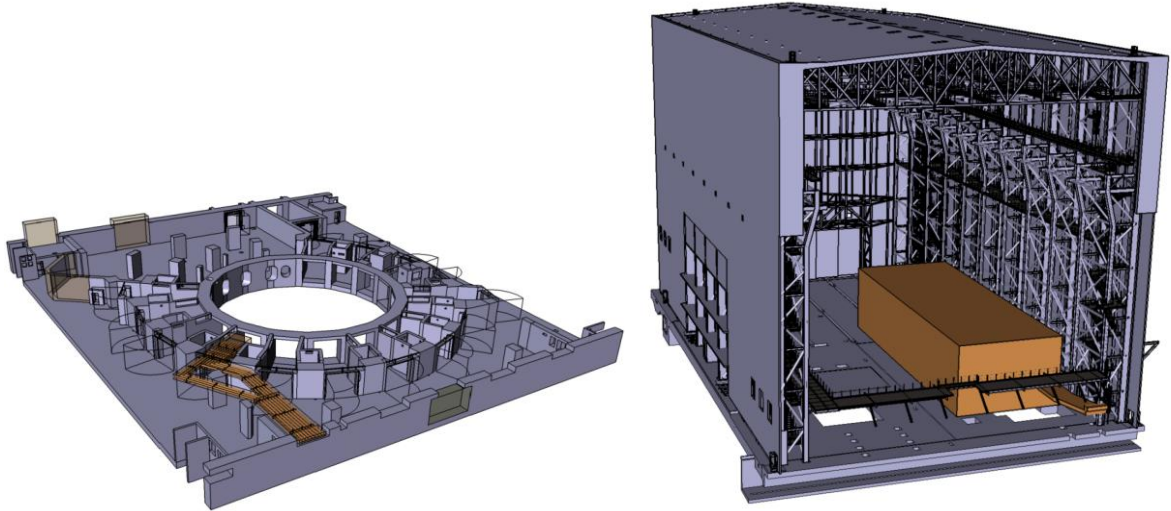

*FIG. 4. Space allocation for the Lower Hybrid system. Left: LH transmission lines in the tokamak building to equatorial port plug t#11. Right: RF sources and HV supplies location in the assembly hall.*

One of the main critical components of a LH system is the RF windows, which separate the tokamak vacuum from the pressurized transmission lines. In ITER, RF windows forms part of the vacuum containment boundary and must fit demanding specifications and qualifications. They should be bakeable to 250°C and be of a double window design, unless permanently installed behind an ultra-high vacuum isolating valve [33]. On the contrary to certain LH antenna designs, the RF windows cannot be located close to the plasma, because of the high heat loads and neutron fluxes expected in ITER. Thus they should be located at the rear of the antenna, which imposes to work at higher RF power density. As a high RF power device, the interspace pressure between two consecutive windows needs to be monitored continuously and interlocked with the power system. Moreover, the windows assembly must comply with the ITER Remote Handling Code of Practice.

In order to address without delay these challenges, a 5 GHz 500 kW/5 s Beryllium-Oxide (BeO) window prototype is under manufacturing. This model is a pill-box window, i.e. a sudden transition from a rectangular waveguide to a circular waveguide in which the ceramic is located. The RF dielectric constant and losses of the BeO ceramic have been measured to be $\varepsilon_r$=6.36 and $\tan(\delta)$=4.5e-4 at 5 GHz. From these results the ceramic width and the cylindrical guide dimensions have been calculated to minimize both the reflected power and the integrated power losses in the ceramic volume. The rectangular waveguides are kept cut-off to higher modes than the fundamental one in order to avoid spurious modes in the transmission line, excited by the rectangular to circular transition. An appropriate TiN coating on the vacuum-side of the ceramic is applied to reduce the secondary electron emission. This prototype will be tested at high power in collaboration with the National Fusion Research Institute (NFRI) in Korea. These tests will allow validating both the RF and the mechanical design for short pulses at high power, before pursuing a CW high power prototype, compatible with the full ITER water cooling, structural and remote handling requirements.

### 4. Conclusion

The results obtained with the ITER relevant 3.7 GHz LH system in Tore Supra represent a significant milestone on the way towards future integration of a LH launcher on ITER. The experience gained during the LH PAM launcher manufacturing allows us to be confident concerning the realization of an ITER LH launcher. The Tore Supra 700 kW/CW 3.7 GHz klystrons have been validated in the test bed at 620 kW/CW on VSWR=1.4, at a power level required for ITER. The sixteen klystrons RF plant, the water cooled pressurized transmission lines and LH launchers are unique tools to develop and test reliable interlocks systems, closely integrated to the main CODAC system. The know-how acquired through the use of the whole Tore Supra CW LH system through the last experimental campaigns, is of the order of the future technical challenges that ITER RF systems will encounter. These technical advances are now complemented by more and more realistic LH simulation tools. The current drive performance can be predicted in identified ranges of plasmas with good accuracy through the use of integrated LH tools, from the RF antenna to the current deposition in the plasma and the associated synthetic diagnostics.

The CEA/IRFM is currently conducting a R&D effort towards the validation of such RF devices. In 2011, a 5 GHz mode converter has been successfully tested at low power, thus validating the RF design. In 2012, a 500 kW BeO ceramic window has been developed and tested at low power. High power tests will be conducted in association with the NFRI/KSTAR LH team.

**Acknowledgment**: Part of this work, supported by the European Communities under the contract of Association between EURATOM and CEA, was carried out within the framework of the European Fusion Development Agreement. The views and opinions expressed herein do not necessarily reflect those of the European Commission.